\documentclass[twocolumn,prl,aps,10pt]{revtex4-1}

\usepackage{amsfonts,amssymb}
\usepackage[sumlimits,intlimits]{amsmath}
\usepackage{graphics}
\usepackage{graphicx}
\usepackage{mathrsfs}
\usepackage{verbatim}
\usepackage{hyperref}    
\usepackage{bm}          

\pdfoutput=1

\newcommand{\be}{\begin{equation}}
\newcommand{\ee}{\end{equation}}

\begin{document}

\title{Coulomb gap triptych in a periodic array of metal nanocrystals}

\date{\today}

\author{Tianran Chen}
\author{Brian Skinner}
\author{B. I. Shklovskii}
\affiliation{Fine Theoretical Physics Institute, University of Minnesota, Minneapolis, MN 55455, USA}

\begin{abstract}

The Coulomb gap in the single particle density of states (DOS) is a universal consequence of electron-electron interaction in disordered systems with localized electron states. Here we show that in arrays of monodisperse metallic nanocrystals, there is not one but three identical adjacent Coulomb gaps, which together form a structure that we call a ``Coulomb gap triptych."  We calculate the DOS and the conductivity in two- and three-dimensional arrays using a computer simulation. Unlike in the conventional Coulomb glass models, in nanocrystal arrays the DOS has a fixed width in the limit of large disorder.  The Coulomb gap triptych can be studied via tunneling experiments.

\end{abstract}
\maketitle

Granular metals and arrays of metallic nanocrystals (NCs) represent interesting composite systems, wherein the unique properties of individual NCs are combined with collective, correlation-driven effects between NCs to produce novel material properties \cite{Talapin2010poc, Moreira2011ect}.  One of the most important of these properties is the electron conductivity, which proceeds by electron tunneling, or ``hopping", between NCs through the insulating gaps which separate them.  In relatively dense NC arrays, electron conduction can occur both through nearest-neighbor hopping and through ``cotunneling" of electrons between distant NCs via a chain of intermediate virtual states \cite{Moreira2011ect}.  
In the presence of some disorder, the latter mechanism dominates at low temperatures, where the length of hops grows to optimize the conductivity.  This transport mechanism was introduced by Mott \cite{Mott1968cig} and is called variable range hopping (VRH).  When the Coulomb interaction between localized electrons is taken into account, it can be shown that at sufficiently low temperature VRH conductivity obeys the Efros-Shklovskii (ES) law \cite{Efros1975cga}:
\be 
\sigma = \sigma_0 \exp \left[ -\left( T_{ES}/T \right)^{1/2} \right],
\label{eq:ES}
\ee 
where $\sigma_0$ is a constant (or a weak, power-law function of temperature) and $T_{ES}$ is a characteristic temperature.  Eq.\ (\ref{eq:ES}) has been observed in a number of granular metal systems at low temperature (see Refs.\ \cite{Beloborodov2007ges} and references therein).  In these systems, as in lightly-doped semiconductors and other ``Coulomb glasses", ES conductivity can be seen as the result of a vanishing single-particle density of states (DOS) at the Fermi level $\mu$.  This vanishing DOS is the consequence of a very general stability criterion of the ground state \cite{Efros1984epo}, and it implies that in a system of $d$ dimensions the DOS $g(E)$ satisfies
\be 
g(E) < \frac{A_d}{e^{2d}} |E|^{d-1}.
\label{eq:Coulombgap}
\ee 
Here, $A_d$ is some numerical constant of order unity, $E$ is the electron energy relative to the Fermi level, and $e$ is the electron charge.  Eq.\ (\ref{eq:Coulombgap}) is called the ``Coulomb gap."

\begin{figure}[htb!]
\centering
\includegraphics[width=0.5 \textwidth]{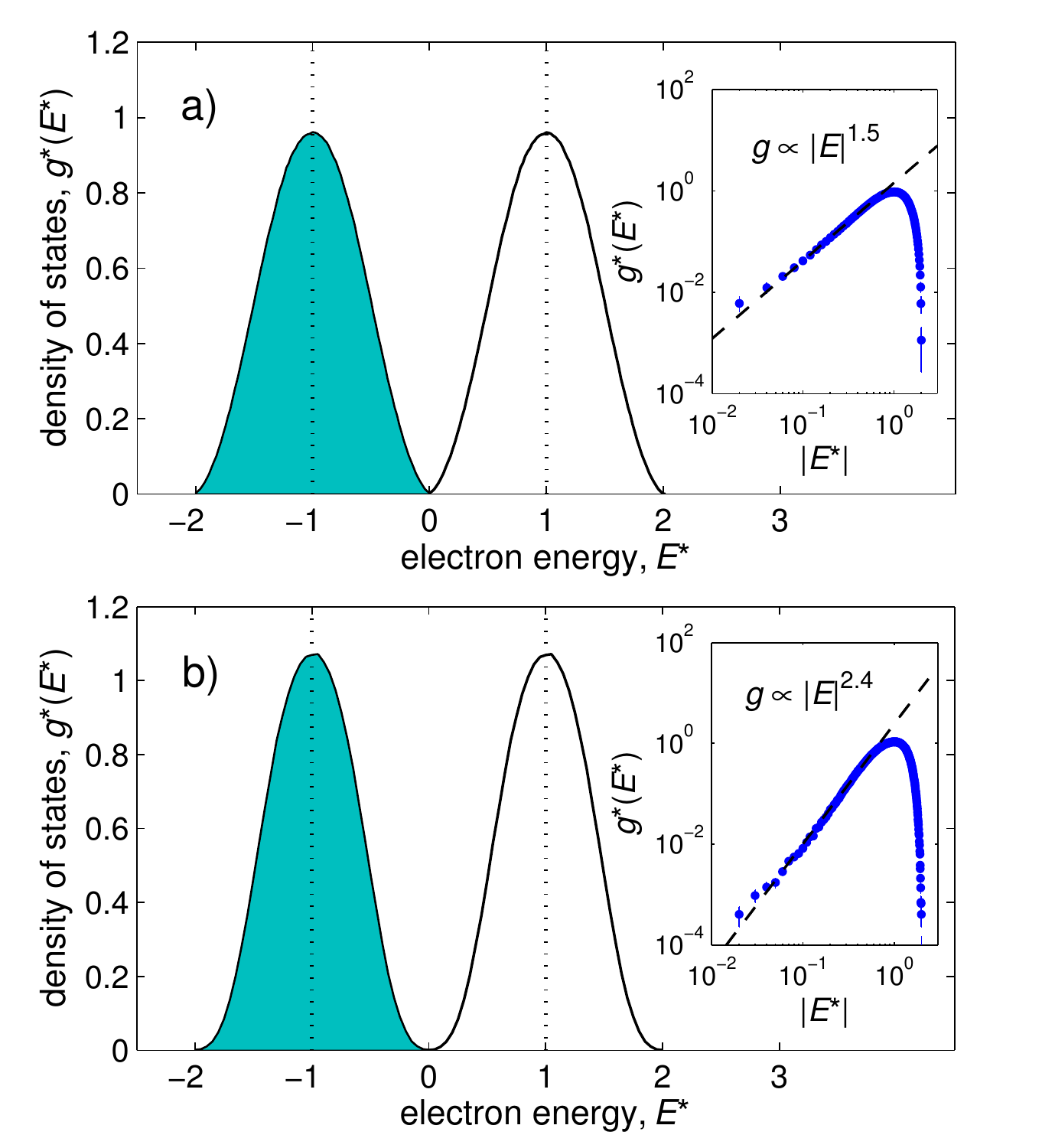}
\caption{(Color online) The DOS of a regular array of monodisperse NCs, where $E^* = E/(e^2/2C_0)$ is the dimensionless single-particle energy and $g^*(E^*) = (e^2 D^{d}/2C_0) g(E^*)$ is the dimensionless DOS, where $D$ is the NC diameter.  Here, the results are shown from a computer simulation of a) a 2d square lattice and b) a 3d cubic lattice.  The shaded area shows filled electron states, and the empty area indicates empty states.  In addition to electron--hole symmetry, the two peaks of the DOS have a mirror symmetry across $E^* = \pm 1$, respectively (dotted lines).  This symmetry creates from the central Coulomb gap two additional half-gaps at $E^* = \pm 2$, resulting in a ``Coulomb gap triptych."  Insets show the DOS near the Fermi level $E^* = 0$ in log-log scale.
} \label{fig:DOS}
\end{figure}

In this letter we report an additional striking feature of the DOS in periodic arrays of monodisperse metal NCs surrounded by random impurity charges.  Namely, we show that the Coulomb gap at $E = 0$ necessarily implies the existence of additional, identical Coulomb gaps at energies $E = \pm e^2/C_0$, where $C_0$ is the self-capacitance of each NC.  This result is shown in Fig.\ \ref{fig:DOS}.  

In Fig.\ \ref{fig:DOS} one can see that to the right of the Fermi level, at $E > 0$, the DOS curve $g(E)$ becomes reflected across the vertical line $E = e^2/2C_0$ into the interval $e^2/2C_0 < E < e^2/C_0$.  Similarly, on the left side of the Fermi level the portion of the $g(E)$ curve corresponding to $-e^2/2C_0 < E < 0$ becomes reflected across the vertical line $E = -e^2/2C_0$.  We refer to this characteristic shape as a ``Coulomb gap triptych."  As we explain below, its structure is the result of an additional symmetry in the system that arises as a result of the discrete charging spectrum of individual NCs.  In this way the Coulomb gap triptych represents a bridge between the concepts of the Coulomb gap and the Coulomb blockade.

Experimentally, regular arrays of metal NCs can now be reliably synthesized with diameter $D$ in the range 3--7 nm and with size dispersion less than 5\% \cite{Beloborodov2007ges, Talapin2010poc, Moreira2011ect}.  For such small NCs, the self-capacitance $C_0$ is also small: $C_0 = \kappa D/2$, where $\kappa$ is the effective dielectric constant of the array, given approximately by the Maxwell-Garnett formula \cite{Maxwell1891tea, Chen2011cci}.  Correspondingly, the Coulomb self-energy $q^2/2C_0$ of an NC with charge $q$ plays a large and important role in electron transport.  To see this, one can imagine a hypothetical NC array with no disorder.  In such an array, in the ground state all NCs are neutral and electron conduction requires the thermal excitation of positive-negative NC pairs.  Thus, the conductivity is activated with an activation energy $e^2/2C_0$.  For nanometer-sized NCs, this activation energy can easily exceed the thermal energy $k_BT$.  

In the presence of some finite charge disorder, however, the fluctuating Coulomb potential can produce charging of NCs in the ground state and thus lead to a Coulomb gap in the DOS and to ES conductivity.  To show how this happens, in this letter we adopt the following simplified model.  We assume that identical, spherical, metallic NCs reside in a regular $d$-dimensional square lattice with lattice constant $D'$, and that impurity charges $\pm e$ are embedded in the insulator (oxide) between NCs.  Such impurity charges can be thought to effectively create a fractional donor charge $Q_i$ that resides on each NC $i$, for reasons that are explained below.  The net charge of the NC can then be written as $q_i = Q_i - en_i$, where $n_i$ is the integer number of electrons that reside on the NC relative to its neutral state ($n_i$ can be positive or negative).  Given this model, the Hamiltonian for the system is
\be
H = \sum_i \frac{(Q_i - en_i)^2 }{2 C_0} 
+ \sum_{\langle i,j \rangle } C^{-1}_{ij} (Q_i - en_i)(Q_j - en_j).
\label{eq:H}
\ee 
Here, the first term describes the Coulomb self-energy of each NC and the second term describes the interaction between charged NCs.  The coefficient $C^{-1}_{ij}$ is the inverse of the matrix of electrostatic induction $C_{ij}$.  This Hamiltonian has been also been proposed as a model for arrays of large semiconductor NCs \cite{Skinner2012toh}.

Because of the presence of the impurity charges, electrons become redistributed among NCs from their neutral state in order to screen the disorder Coulomb potential.
In order to calculate the DOS and conductivity we first attempt to find numerically the set of electron occupation numbers $\{n_i\}$ that minimizes the Hamiltonian.
In the numerical simulations that we describe below, we make the approximations that $C_0 = \kappa D/2$ and $C^{-1}_{ij} = 1/\kappa r_{ij}$.  These approximations do not effect our main conclusions, as we explain below. 

\begin{figure}[htb!]
\centering
\includegraphics[width=0.45 \textwidth]{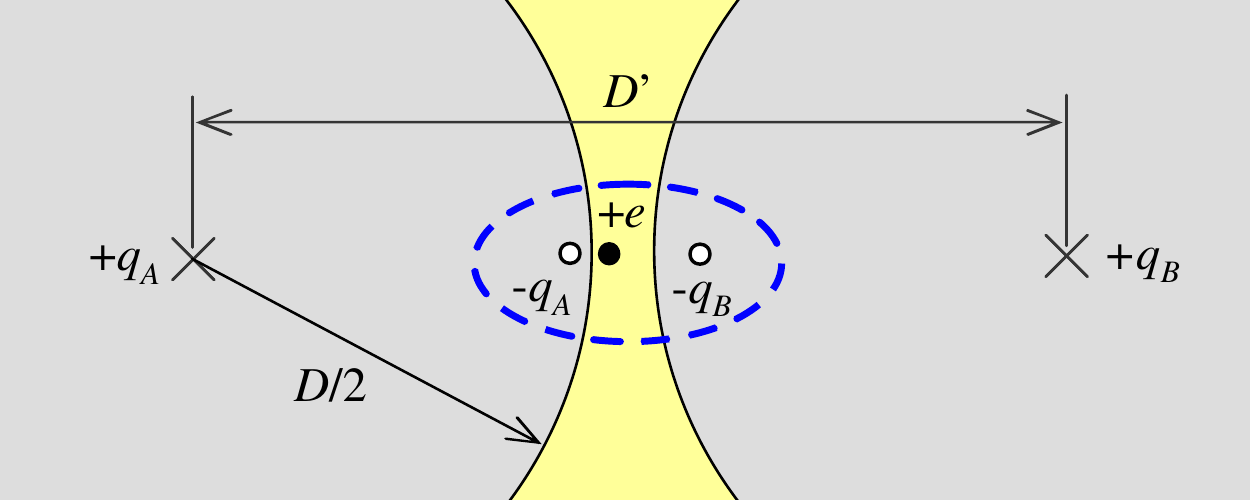}
\caption{(Color online) A schematic depiction of the fractionalization of a charged impurity (small black circle) between NCs (large gray circles).  The positive impurity induces negative image charges (white circles) in nearby metal surfaces and is effectively neutralized, while equal and opposite positive images are conveyed to the center of the NC ($\times$'s).} \label{fig:fracschematic}
\end{figure}

The model of fractional donor charges $Q_i$ was first put forward in Ref.\ \cite{Zhang2004dos}; here its justification is briefly repeated.  When an impurity charge, say with charge $+e$, is located close to the point of contact between two NCs, labeled A and B, it induces negative image charges $-q_A$ and $-q_B$ in the surfaces of NCs A and B, respectively.  This is shown schematically in Fig.\ \ref{fig:fracschematic}.  In order to maintain overall neutrality of the NCs, an equal and opposite image charge appears at the center of each NC: $+q_A$ and $+q_B$.  (These ``image charges at the center" represent a uniform electronic charge at the NC surface.)  The values of $q_A$ and $q_B$ are such that together the image charges $-q_A$ and $-q_B$ neutralize the donor charge: $q_A + q_B = e$.  Their respective magnitudes are determined by the distance between the impurity and each NC surface.  For example, if the impurity sits exactly along the line connecting the centers of NCs A and B and if the gap $w = D' - D$ between NCs satisfies $w \ll D$, then $q_A x_B = q_B x_A$, where $x_A$ and $x_B$ are the distances between the impurity and the surface of NCs A and B, respectively.  Since the impurity charges and the image charges $-q_A$, $-q_B$ together form a compact, neutral arrangement, the net effect of the impurity charge is to produce ``fractionalized" donor charges, such that $+q_A$ is relayed to the center of NC A and $+q_B$ is relayed to the center of B.

In this way, each NC $i$ can be said to have a fractional donor charge $Q_i$, which is equal to the sum total of the fractionalized charges donated by individual impurities around it.  In the limit where there are very many impurity charges surrounding each NC, one can think that the random variable $Q_i$ is Gaussian-distributed with some standard deviation larger than $e$.  In fact, however, in such cases one can effectively adopt a much simpler model, in which the value of $Q_i$ is chosen randomly from the uniform distribution $Q_i \in [-e/2, +e/2]$.  To see why this model is valid, consider that each NC minimizes its Coulomb self-energy by minimizing the magnitude of its net charge, $|Q_i - en_i|$.  Since $n_i$ can take any integer value, it is generally true that in the ground state $-e/2 \leq Q_i - en_i \leq e/2$.  In other words, each NC can effectively adjust to the presence of an arbitrarily strong charge disorder by changing its electron number $n_i$ (say, by drawing electrons from the voltage source) so that its net charge acquires a magnitude smaller than $e/2$.  This has important implications for the disorder-dependence of conductivity, as we show below.

Given the ground state configuration for a particular system, defined by the set of electron occupation numbers $\{n_i\}$, one can determine the energy of the highest filled electron level, $E_i^{(f)}$, and the lowest empty electron level, $E_i^{(e)}$, at each NC $i$.  Specifically:
\begin{eqnarray}
E_i^{(f)} = \frac{2 e^2 n_i - 2 Q_i e - e^2}{2 C_0} - e\sum_{j \neq i} C^{-1}_{ij}(Q_j - en_j),
\label{eq:enf} \\
E_i^{(e)} = \frac{2 e^2 n_i - 2 Q_i e + e^2}{2 C_0} - e\sum_{j \neq i} C^{-1}_{ij}(Q_j - en_j).
\label{eq:ene}
\end{eqnarray}
These energies are defined so that the Fermi level $\mu = 0$, and in the ground state $E_i^{(f)} < 0$ and $E_i^{(e)} > 0$ for all $i$.  The single particle DOS $g(E)$ is defined by making a histogram of the energy values $E_i^{(f)}$ and $E_i^{(e)}$.  Higher and lower electron energy states are ignored in this work, as they play no role in conductivity at $k_BT \ll e^2/C_0$.

In order to evaluate numerically the DOS, we use a computer simulation to search for the ground state arrangement of electrons, $\{n_i\}$, in a finite array of NCs.  For simplicity, we set the lattice constant $D' = D$; this corresponds to the limit where the gap $w$ between NCs is very thin while the tunneling transparency of the barrier between them remains much smaller than unity.   In our simulation we search for the ground state by looping over all NC pairs $i, j$ and attempting to move one electron from $i$ to $j$.  If the move lowers the Hamiltonian $H$, then it is accepted, otherwise it is rejected.  Equivalently, one can say that for all $i$, $j$ we check that the ES ground state criterion is satisfied:
\be 
E_{j}^{(e)} - E_i^{(f)} - e^2 C^{-1}_{ij} > 0.
\label{eq:EScrit}
\ee
It should be noted that this procedure does not in general find the exact ground state, but only a ``pseudo-ground state" that is stable with respect to single-electron transfers.  In principle, the system energy can be lowered further by some simultaneous multi-electron transfers.  Such processes are generally seen to have only a relatively weak effect on the DOS \cite{Mobius1992cgi, Efros2011cgi} that slightly deepens the Coulomb gap near the Fermi level.

The resulting DOS is shown in Fig.\ \ref{fig:DOS}a for a two-dimensional (2d) simulated system of size $100 \times 100$ lattice sites and in Fig.\ \ref{fig:DOS}b for a three-dimensional (3d) system of size $25 \times 25 \times 25$.  Electron energies are plotted in the dimensionless form $E^* = E/(e^2/2C_0)$ and the DOS is plotted in the dimensionless form $g^*(E^*) = (e^2 D^{d}/2C_0) g(E^*)$.  The insets to these figures show a log-log plot of the DOS near $E = 0$, which suggests that in 2d the DOS follows $g_\text{2d}(E) \propto E^{1.5}$ at small energies and in 3d $g_\text{3d}(E) \propto E^{2.4}$.  These exponents are somewhat larger than the theoretical ones given in Eq.\ (\ref{eq:ES}), so that apparently the ES bound is not saturated.  This is similar to what happens in the Efros model  of the Coulomb glass \cite{Efros1976cgi} at disorder strength $A = 1$ \cite{Mobius1992cgi}.  The results of Fig.\ \ref{fig:DOS} are generated using a uniform distribution $Q_i \in [-e/2, e/2]$ for the fractional charge.  If one instead takes $Q_i$ to be Gaussian-distributed with a standard deviation $< 3e$, the resulting DOS is everywhere equal to that of Fig.\ \ref{fig:DOS} to within $0.6\%$.  

Fig.\ \ref{fig:DOS} also highlights the striking additional symmetry in the DOS in both 2d and 3d, as compared to the DOS in the conventional Coulomb glass problem \cite{Efros1984epo, Mobius1992cgi}.  Namely, each peak in the DOS is symmetric with respect to reflections about $E^* = \pm 1$, so that the DOS has identical, repeated Coulomb gaps at $E^* = \pm 2$.  The origin of these additional Coulomb gaps can be understood by noting a particular symmetry in the Hamiltonian that is reflected in the filled and empty state energies, $E_i^{(f)}$ and $E_i^{(e)}$.  Namely, by subtracting Eqs.\ (\ref{eq:enf}) and (\ref{eq:ene}) one can show that 
\be 
E_i^{*(e)} = E_i^{*(f)} + 2
\label{eq:eplus2}
\ee
for all $i$.  Thus, all NCs contribute to the DOS two energy levels -- one filled, one empty -- separated by $e^2/C_0$.  This implies that as the density of states collapses at $E$ very close to zero (the Coulomb gap), the density of states must also collapse as $E^*$ approaches $\pm 2$ in identical fashion.  That is, the ES stability criterion of Eq.\ (\ref{eq:EScrit}) places constraints both on the DOS near $E = 0$ and on the DOS near $E = \pm e^2/C_0$.

One can also note that states with $E_i^{*(f)} < -2 $ or $E_i^{*(e)} > 2$ are prohibited, since by Eq.\ (\ref{eq:eplus2}) these would imply that some NC has $E_i^{(e)} < 0$ or $E_i^{(f)} > 0$.  Thus, $g(E)$ is strictly zero at $|E^*| > 2$.  This is a markedly different situation than in the conventional Efros model \cite{Efros1976cgi}, where the width of the DOS reflects the characteristic strength of the disorder.  In the present problem, for large enough disorder the DOS has a saturated width $e^2/C_0$.  This saturation occurs because the number of electrons $n$ at each site can adjust to screen an arbitrarily large Coulomb disorder.  Thus, one can expect that at large disorder the conductivity also becomes independent of disorder strength.

In order to evaluate the conductivity directly, we employ the approach of the Miller-Abrahams network \cite{Miller1960ica}, in which each pair $ij$ of NCs is said to be connected by some equivalent resistance $R_{ij}$.  The value of $R_{ij}$ increases exponentially with the distance $r_{ij}$ between NCs and the activation energy $\Delta E_{ij}$ required for electron hopping between $i$ and $j$ according to $R_{ij} \propto \exp[2 r_{ij}/\xi + \Delta E_{ij}/k_BT]$, where $\xi$ is the electron localization length \cite{Zhang2004dos} and the value of $\Delta E_{ij}$ is determined by the ground state energies $\{E_i^{(f)}\}$ and $\{E_i^{(e)}\}$ \cite{Skinner2012toh}.  The resistance of the system as a whole can be found using a percolation approach.  Specifically, we find the minimum value $R_c$ such that if all resistances $R_{ij}$ with $R_{ij} < R_c$ are left intact while others are eliminated (replaced with $R = \infty$), then there exists a percolation pathway connecting opposite faces of the simulation volume.  The conductivity of the system $\sigma$ is equated with $1/(R_c D^{d-2})$.

\begin{figure}[htb!]
\centering
\includegraphics[width=0.5 \textwidth]{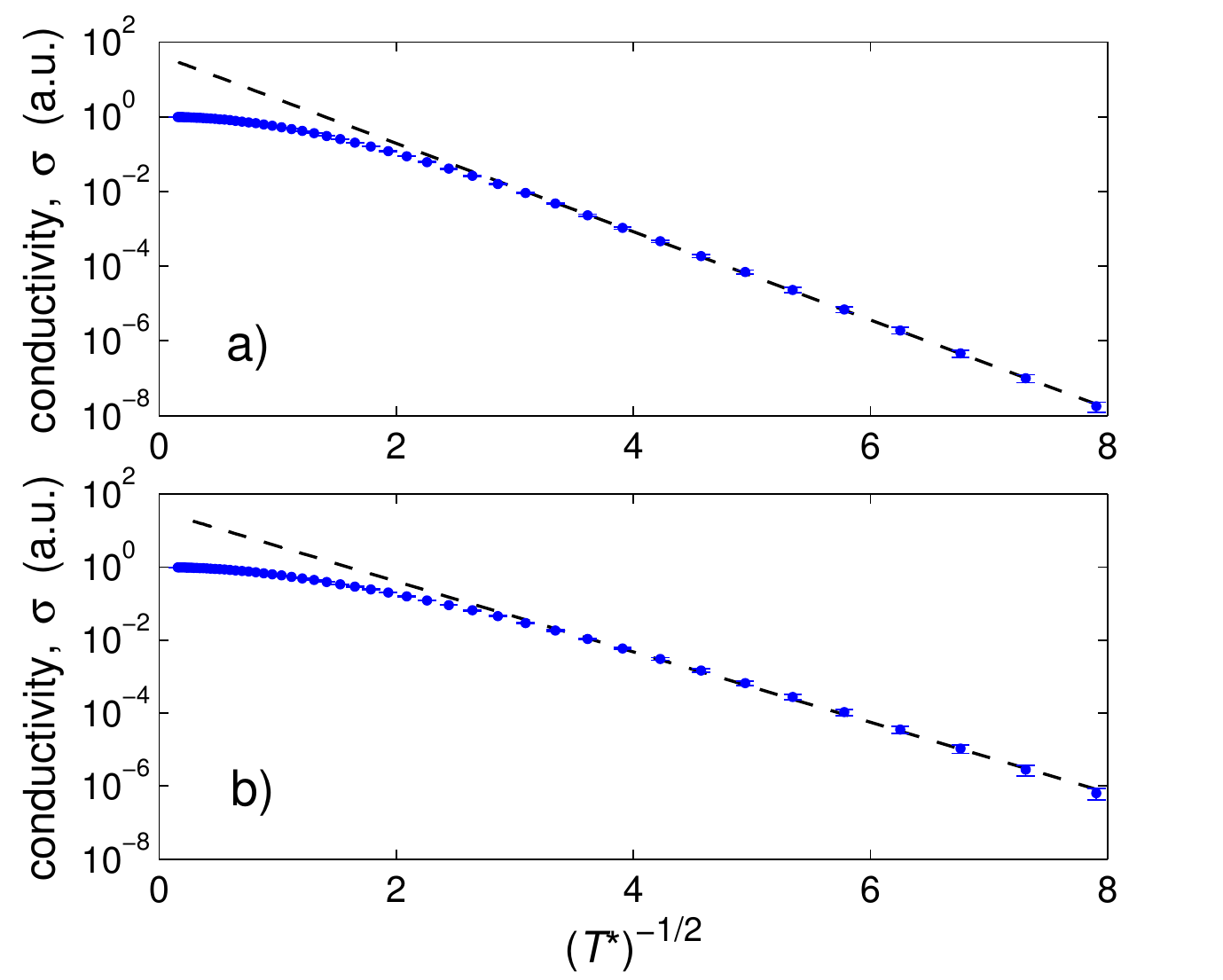}
\caption{(Color online) The temperature dependence of the conductivity in a) 2d and b) 3d.  In both cases, the conductivity follows the ES law [Eq.\ (\ref{eq:ES})] at small temperatures, $T^* \ll 1$, as shown by the dashed lines.} \label{fig:sigma}
\end{figure}

Our results for the conductivity are shown in Fig.\ \ref{fig:sigma}, plotted as a function of the dimensionless temperature $T^* = 4 D C_0 k_BT/(e^2 \xi)$ raised to the power $-1/2$.  The results indicate that the conductivity is well-described by the ES law of Eq.\ (\ref{eq:ES}) at relatively small temperatures $T^* \lesssim 1$, both in 2d and 3d
\footnote{In fact, if one repeats the original ES derivation \cite{Efros1975cga} using the DOS shown in Fig.\ \ref{fig:DOS}, one arrives at a slightly different temperature dependence $\ln \sigma \propto T^{-\gamma}$ at low temperature, with $\gamma \approx 0.56$ in 2d and $\gamma \approx 0.53$ in 3d.  Due to finite size limitations, our conductivity data (Fig.\ \ref{fig:sigma}) cannot discriminate between these exponents and $\gamma = 1/2$.}.  
This behavior is consistent with the prominent Coulomb gaps seen in Fig.\ \ref{fig:DOS}. In both 2d and 3d, replacing the uniform distribution of $Q_i$ with a distribution with larger variance --- for example, by taking $Q_i$ as the sum of three or more independent fractional charges --- did not affect the conductivity to within our numerical accuracy.  This insensitivity to the disorder strength stands in contrast to the Efros model \cite{Efros1976cgi}, where large disorder widens the DOS, so that ES conductivity exists only when the temperature is sufficiently small that electron hops are confined to within the parametrically narrow window of energies in which $g(E)$ is constrained by the Coulomb gap \cite{Efros1984epo}.  On the contrary, in arrays of monodisperse metallic NCs the DOS becomes essentially independent of disorder strength, so that even at large disorder the Coulomb gap plays a prominent role and the conductivity follows the ES law.

The triptych structure of the DOS should have observable consequences for a number of experiments on metal NC arrays.  It is possible, for example, that the DOS can be probed directly by tunneling experiments, similar to the ones that have directly observed the Coulomb gap in doped semiconductors \cite{Massey1995doo}.  For systems with a finite dispersion $\delta C$ in the NC self-capacitance, the repeated Coulomb gaps will be smeared over some finite energy interval rather than collapsing to zero exactly at $E^* = \pm 2$.  One can simulate this behavior numerically by adding a stochastic spatial variation to $C_0$.  Our simulations suggest that for root mean square deviation $\delta C \ll C_0$, $g(E^* = \pm 2)/g(E^* = \pm 1) \approx 3(\delta C/C_0)^2$.  This implies that for a system with $5\%$ dispersion in the NC diameter, the collapse of the DOS at $E^* = \pm 2$ is complete to within $1\%$, and the resulting $g^*(E^*)$ curve would not be distinguishable from that of Fig.\ \ref{fig:DOS} if added to the plot.

The authors would like to thank E. Aydil, A. L. Efros, A. Frydman, Yu. M. Galperin, M. Goethe, A. Kamenev, U. Kortshagen, M. Muller, A M\"{o}bius, M. Palassini, and L. Wienkes for helpful discussions.
This work was supported primarily by the MRSEC Program of the National Science Foundation under Award Number DMR-0819885.  T. Chen was partially supported by the FTPI.

\bibliography{metal_paper}

\begin{thebibliography}{16}%
\makeatletter
\providecommand \@ifxundefined [1]{%
 \@ifx{#1\undefined}
}%
\providecommand \@ifnum [1]{%
 \ifnum #1\expandafter \@firstoftwo
 \else \expandafter \@secondoftwo
 \fi
}%
\providecommand \@ifx [1]{%
 \ifx #1\expandafter \@firstoftwo
 \else \expandafter \@secondoftwo
 \fi
}%
\providecommand \natexlab [1]{#1}%
\providecommand \enquote  [1]{``#1''}%
\providecommand \bibnamefont  [1]{#1}%
\providecommand \bibfnamefont [1]{#1}%
\providecommand \citenamefont [1]{#1}%
\providecommand \href@noop [0]{\@secondoftwo}%
\providecommand \href [0]{\begingroup \@sanitize@url \@href}%
\providecommand \@href[1]{\@@startlink{#1}\@@href}%
\providecommand \@@href[1]{\endgroup#1\@@endlink}%
\providecommand \@sanitize@url [0]{\catcode `\\12\catcode `\$12\catcode
  `\&12\catcode `\#12\catcode `\^12\catcode `\_12\catcode `\%12\relax}%
\providecommand \@@startlink[1]{}%
\providecommand \@@endlink[0]{}%
\providecommand \url  [0]{\begingroup\@sanitize@url \@url }%
\providecommand \@url [1]{\endgroup\@href {#1}{\urlprefix }}%
\providecommand \urlprefix  [0]{URL }%
\providecommand \Eprint [0]{\href }%
\providecommand \doibase [0]{http://dx.doi.org/}%
\providecommand \selectlanguage [0]{\@gobble}%
\providecommand \bibinfo  [0]{\@secondoftwo}%
\providecommand \bibfield  [0]{\@secondoftwo}%
\providecommand \translation [1]{[#1]}%
\providecommand \BibitemOpen [0]{}%
\providecommand \bibitemStop [0]{}%
\providecommand \bibitemNoStop [0]{.\EOS\space}%
\providecommand \EOS [0]{\spacefactor3000\relax}%
\providecommand \BibitemShut  [1]{\csname bibitem#1\endcsname}%
\let\auto@bib@innerbib\@empty
\bibitem [{\citenamefont {Talapin}\ \emph {et~al.}(2010)\citenamefont
  {Talapin}, \citenamefont {Lee}, \citenamefont {Kovalenko},\ and\
  \citenamefont {Shevchenko}}]{Talapin2010poc}%
  \BibitemOpen
  \bibfield  {author} {\bibinfo {author} {\bibfnamefont {D.~V.}\ \bibnamefont
  {Talapin}}, \bibinfo {author} {\bibfnamefont {J.-S.}\ \bibnamefont {Lee}},
  \bibinfo {author} {\bibfnamefont {M.~V.}\ \bibnamefont {Kovalenko}}, \ and\
  \bibinfo {author} {\bibfnamefont {E.~V.}\ \bibnamefont {Shevchenko}},\ }\href
  {\doibase 10.1021/cr900137k} {\bibfield  {journal} {\bibinfo  {journal}
  {Chemical Reviews}\ }\textbf {\bibinfo {volume} {110}},\ \bibinfo {pages}
  {389} (\bibinfo {year} {2010})}\BibitemShut {NoStop}%
\bibitem [{\citenamefont {Moreira}\ \emph {et~al.}(2011)\citenamefont
  {Moreira}, \citenamefont {Yu}, \citenamefont {Nadal}, \citenamefont
  {Bresson}, \citenamefont {Rosticher}, \citenamefont {Lequeux}, \citenamefont
  {Zimmers},\ and\ \citenamefont {Aubin}}]{Moreira2011ect}%
  \BibitemOpen
  \bibfield  {author} {\bibinfo {author} {\bibfnamefont {H.}~\bibnamefont
  {Moreira}}, \bibinfo {author} {\bibfnamefont {Q.}~\bibnamefont {Yu}},
  \bibinfo {author} {\bibfnamefont {B.}~\bibnamefont {Nadal}}, \bibinfo
  {author} {\bibfnamefont {B.}~\bibnamefont {Bresson}}, \bibinfo {author}
  {\bibfnamefont {M.}~\bibnamefont {Rosticher}}, \bibinfo {author}
  {\bibfnamefont {N.}~\bibnamefont {Lequeux}}, \bibinfo {author} {\bibfnamefont
  {A.}~\bibnamefont {Zimmers}}, \ and\ \bibinfo {author} {\bibfnamefont
  {H.}~\bibnamefont {Aubin}},\ }\href {\doibase 10.1103/PhysRevLett.107.176803}
  {\bibfield  {journal} {\bibinfo  {journal} {Phys. Rev. Lett.}\ }\textbf
  {\bibinfo {volume} {107}},\ \bibinfo {pages} {176803} (\bibinfo {year}
  {2011})}\BibitemShut {NoStop}%
\bibitem [{\citenamefont {Mott}(1968)}]{Mott1968cig}%
  \BibitemOpen
  \bibfield  {author} {\bibinfo {author} {\bibfnamefont {N.}~\bibnamefont
  {Mott}},\ }\href {\doibase 10.1016/0022-3093(68)90002-1} {\bibfield
  {journal} {\bibinfo  {journal} {Journal of Non-Crystalline Solids}\ }\textbf
  {\bibinfo {volume} {1}},\ \bibinfo {pages} {1 } (\bibinfo {year}
  {1968})}\BibitemShut {NoStop}%
\bibitem [{\citenamefont {Efros}\ and\ \citenamefont
  {Shklovskii}(1975)}]{Efros1975cga}%
  \BibitemOpen
  \bibfield  {author} {\bibinfo {author} {\bibfnamefont {A.~L.}\ \bibnamefont
  {Efros}}\ and\ \bibinfo {author} {\bibfnamefont {B.~I.}\ \bibnamefont
  {Shklovskii}},\ }\href {\doibase 10.1088/0022-3719/8/4/003} {\bibfield
  {journal} {\bibinfo  {journal} {J. Phys. C: Solid State Phys.}\ }\textbf
  {\bibinfo {volume} {8}},\ \bibinfo {pages} {L49} (\bibinfo {year}
  {1975})}\BibitemShut {NoStop}%
\bibitem [{\citenamefont {Beloborodov}\ \emph {et~al.}(2007)\citenamefont
  {Beloborodov}, \citenamefont {Lopatin}, \citenamefont {Vinokur},\ and\
  \citenamefont {Efetov}}]{Beloborodov2007ges}%
  \BibitemOpen
  \bibfield  {author} {\bibinfo {author} {\bibfnamefont {I.~S.}\ \bibnamefont
  {Beloborodov}}, \bibinfo {author} {\bibfnamefont {A.~V.}\ \bibnamefont
  {Lopatin}}, \bibinfo {author} {\bibfnamefont {V.~M.}\ \bibnamefont
  {Vinokur}}, \ and\ \bibinfo {author} {\bibfnamefont {K.~B.}\ \bibnamefont
  {Efetov}},\ }\href {\doibase 10.1103/RevModPhys.79.469} {\bibfield  {journal}
  {\bibinfo  {journal} {Rev. Mod. Phys.}\ }\textbf {\bibinfo {volume} {79}},\
  \bibinfo {pages} {469} (\bibinfo {year} {2007})}\BibitemShut {NoStop}%
\bibitem [{\citenamefont {Efros}\ and\ \citenamefont
  {Shklovskii}(1984)}]{Efros1984epo}%
  \BibitemOpen
  \bibfield  {author} {\bibinfo {author} {\bibfnamefont {A.~L.}\ \bibnamefont
  {Efros}}\ and\ \bibinfo {author} {\bibfnamefont {B.~I.}\ \bibnamefont
  {Shklovskii}},\ }\href@noop {} {\emph {\bibinfo {title} {Electronic
  Properties of Doped Semiconductors}}}\ (\bibinfo  {publisher}
  {Springer-Verlag},\ \bibinfo {address} {New York},\ \bibinfo {year}
  {1984})\BibitemShut {NoStop}%
\bibitem [{\citenamefont {Maxwell}(1891)}]{Maxwell1891tea}%
  \BibitemOpen
  \bibfield  {author} {\bibinfo {author} {\bibfnamefont {J.~C.}\ \bibnamefont
  {Maxwell}},\ }\href@noop {} {\emph {\bibinfo {title} {A Treatise on
  Electricity and Magnetism}}},\ \bibinfo {edition} {3rd}\ ed.,\ Vol.~\bibinfo
  {volume} {2}\ (\bibinfo  {publisher} {Clarendon},\ \bibinfo {address}
  {Oxford},\ \bibinfo {year} {1891})\ p.~\bibinfo {pages} {57}\BibitemShut
  {NoStop}%
\bibitem [{\citenamefont {Chen}\ \emph {et~al.}(2011)\citenamefont {Chen},
  \citenamefont {Skinner},\ and\ \citenamefont {Shklovskii}}]{Chen2011cci}%
  \BibitemOpen
  \bibfield  {author} {\bibinfo {author} {\bibfnamefont {T.}~\bibnamefont
  {Chen}}, \bibinfo {author} {\bibfnamefont {B.}~\bibnamefont {Skinner}}, \
  and\ \bibinfo {author} {\bibfnamefont {B.~I.}\ \bibnamefont {Shklovskii}},\
  }\href {\doibase 10.1103/PhysRevB.84.245304} {\bibfield  {journal} {\bibinfo
  {journal} {Phys. Rev. B}\ }\textbf {\bibinfo {volume} {84}},\ \bibinfo
  {pages} {245304} (\bibinfo {year} {2011})}\BibitemShut {NoStop}%
\bibitem [{\citenamefont {{Skinner}}\ \emph {et~al.}(2012)\citenamefont
  {{Skinner}}, \citenamefont {{Chen}},\ and\ \citenamefont
  {{Shklovskii}}}]{Skinner2012toh}%
  \BibitemOpen
  \bibfield  {author} {\bibinfo {author} {\bibfnamefont {B.}~\bibnamefont
  {{Skinner}}}, \bibinfo {author} {\bibfnamefont {T.}~\bibnamefont {{Chen}}}, \
  and\ \bibinfo {author} {\bibfnamefont {B.~I.}\ \bibnamefont {{Shklovskii}}},\
  }\href@noop {} {\bibfield  {journal} {\bibinfo  {journal} {ArXiv e-prints}\ }
  (\bibinfo {year} {2012})},\ \Eprint {http://arxiv.org/abs/1203.3889}
  {arXiv:1203.3889 [cond-mat.mes-hall]} \BibitemShut {NoStop}%
\bibitem [{\citenamefont {Zhang}\ and\ \citenamefont
  {Shklovskii}(2004)}]{Zhang2004dos}%
  \BibitemOpen
  \bibfield  {author} {\bibinfo {author} {\bibfnamefont {J.}~\bibnamefont
  {Zhang}}\ and\ \bibinfo {author} {\bibfnamefont {B.~I.}\ \bibnamefont
  {Shklovskii}},\ }\href {\doibase 10.1103/PhysRevB.70.115317} {\bibfield
  {journal} {\bibinfo  {journal} {Phys. Rev. B}\ }\textbf {\bibinfo {volume}
  {70}},\ \bibinfo {pages} {115317} (\bibinfo {year} {2004})}\BibitemShut
  {NoStop}%
\bibitem [{\citenamefont {M\"obius}\ \emph {et~al.}(1992)\citenamefont
  {M\"obius}, \citenamefont {Richter},\ and\ \citenamefont
  {Drittler}}]{Mobius1992cgi}%
  \BibitemOpen
  \bibfield  {author} {\bibinfo {author} {\bibfnamefont {A.}~\bibnamefont
  {M\"obius}}, \bibinfo {author} {\bibfnamefont {M.}~\bibnamefont {Richter}}, \
  and\ \bibinfo {author} {\bibfnamefont {B.}~\bibnamefont {Drittler}},\ }\href
  {\doibase 10.1103/PhysRevB.45.11568} {\bibfield  {journal} {\bibinfo
  {journal} {Phys. Rev. B}\ }\textbf {\bibinfo {volume} {45}},\ \bibinfo
  {pages} {11568} (\bibinfo {year} {1992})}\BibitemShut {NoStop}%
\bibitem [{\citenamefont {Efros}\ \emph {et~al.}(2011)\citenamefont {Efros},
  \citenamefont {Skinner},\ and\ \citenamefont {Shklovskii}}]{Efros2011cgi}%
  \BibitemOpen
  \bibfield  {author} {\bibinfo {author} {\bibfnamefont {A.~L.}\ \bibnamefont
  {Efros}}, \bibinfo {author} {\bibfnamefont {B.}~\bibnamefont {Skinner}}, \
  and\ \bibinfo {author} {\bibfnamefont {B.~I.}\ \bibnamefont {Shklovskii}},\
  }\href {\doibase 10.1103/PhysRevB.84.064204} {\bibfield  {journal} {\bibinfo
  {journal} {Phys. Rev. B}\ }\textbf {\bibinfo {volume} {84}},\ \bibinfo
  {pages} {064204} (\bibinfo {year} {2011})}\BibitemShut {NoStop}%
\bibitem [{\citenamefont {Efros}(1976)}]{Efros1976cgi}%
  \BibitemOpen
  \bibfield  {author} {\bibinfo {author} {\bibfnamefont {A.~L.}\ \bibnamefont
  {Efros}},\ }\href {http://stacks.iop.org/0022-3719/9/i=11/a=012} {\bibfield
  {journal} {\bibinfo  {journal} {Journal of Physics C: Solid State Physics}\
  }\textbf {\bibinfo {volume} {9}},\ \bibinfo {pages} {2021} (\bibinfo {year}
  {1976})}\BibitemShut {NoStop}%
\bibitem [{\citenamefont {Miller}\ and\ \citenamefont
  {Abrahams}(1960)}]{Miller1960ica}%
  \BibitemOpen
  \bibfield  {author} {\bibinfo {author} {\bibfnamefont {A.}~\bibnamefont
  {Miller}}\ and\ \bibinfo {author} {\bibfnamefont {E.}~\bibnamefont
  {Abrahams}},\ }\href {\doibase 10.1103/PhysRev.120.745} {\bibfield  {journal}
  {\bibinfo  {journal} {Phys. Rev.}\ }\textbf {\bibinfo {volume} {120}},\
  \bibinfo {pages} {745} (\bibinfo {year} {1960})}\BibitemShut {NoStop}%
\bibitem [{Note1()}]{Note1}%
  \BibitemOpen
  \bibinfo {note} {In fact, if one repeats the original ES derivation \cite
  {Efros1975cga} using the DOS shown in Fig.\ \ref {fig:DOS}, one arrives at a
  slightly different temperature dependence $\protect \qopname \relax
  o{ln}\sigma \propto T^{-\gamma }$ at low temperature, with $\gamma \approx
  0.56$ in 2d and $\gamma \approx 0.53$ in 3d. Due to finite size limitations,
  our conductivity data (Fig.\ \ref {fig:sigma}) cannot discriminate between
  these exponents and $\gamma = 1/2$.}\BibitemShut {Stop}%
\bibitem [{\citenamefont {Massey}\ and\ \citenamefont
  {Lee}(1995)}]{Massey1995doo}%
  \BibitemOpen
  \bibfield  {author} {\bibinfo {author} {\bibfnamefont {J.~G.}\ \bibnamefont
  {Massey}}\ and\ \bibinfo {author} {\bibfnamefont {M.}~\bibnamefont {Lee}},\
  }\href {\doibase 10.1103/PhysRevLett.75.4266} {\bibfield  {journal} {\bibinfo
   {journal} {Phys. Rev. Lett.}\ }\textbf {\bibinfo {volume} {75}},\ \bibinfo
  {pages} {4266} (\bibinfo {year} {1995})}\BibitemShut {NoStop}%
\end{thebibliography}%
\end{document}